\documentclass{webofc}
\usepackage[varg]{txfonts}   % Web of Conferences font
\usepackage{lipsum}
\usepackage{xcolor}
\usepackage{indentfirst}
\usepackage{mdframed}
\usepackage{color,soul}
\usepackage{array}
\usepackage{wrapfig}

\usepackage{titlesec}
\titlespacing*{\section}{0pt}{0.5\baselineskip}{0.4\baselineskip}

\usepackage{aasmacros}
\begin{document}
\title{The European Low Frequency Survey}
%
% subtitle is optionnal
%
\subtitle{Observing the radio sky to understand the beginning of the Universe}

\author{\lastname{A.~Mennella}\inst{1,2}\fnsep\thanks{\email{aniello.mennella@fisica.unimi.it}} \and
        \lastname{K.~Arnold}\inst{3} \and
        \lastname{S.~Azzoni}\inst{4} \and
	 \lastname{C.~Baccigalupi}\inst{5} \and
	\lastname{A.~Banday}\inst{6} \and 
	\lastname{R.B.~Barreiro}\inst{7} \and 
	\lastname{D.~Barron}\inst{8} \and 
	\lastname{M.~Bersanelli}\inst{1,2} \and
	\lastname{S.~Casey}\inst{3} \and
	\lastname{L.~Colombo} \inst{1,2} \and
	\lastname{E.~de la Hoz}\inst{9,10} \and
       \lastname{C.~Franceschet}\inst{1,2} \and
       \lastname{M.E.~Jones} \inst{11} \and
       \lastname{R.T.~Genova-Santos} \inst{12,13} \and
       \lastname{R.J.~Hoyland} \inst{12,13} \and
       \lastname{A.T.~Lee} \inst{10} \and
       \lastname{E.~Martinez-Gonzalez} \inst{7} \and
       \lastname{F.~Montonati} \inst{1} \and
       \lastname{J.A.~Rubi\~{n}o-Martin} \inst{12,13} \and
       \lastname{A.~Taylor} \inst{11} \and
       \lastname{P.~Vielva} \inst{7}
}

\institute{Università degli studi di Milano, Milano, Italy %1
\and
	Istituto Nazionale di Fisica Nucleare (INFN) sezione di Milano, Milano, Italy %2
\and
	University of California San Diego, La Jolla, CA 92093-0021, USA %3
\and
	Princeton University, Princeton, NJ 08544, USA %4
\and
	Scuola Internazionale Superiore di Studi Avanzati (SISSA), Trieste, Italy %5
\and
	Institut de Recherche end Astrophysique et Planétologie (IRAP), Toulouse, France %6
\and
	Instituto de Física de Cantabria (IFCA), Santander, Cantabria, Spain %7
\and
	The University of New Mexico, Albuquerque, NM 87131, USA %8
\and
    Centre Pierre Bin\'etruy (CNRS-UCB), IRL2007, CPB-IN2P3, Berkeley, CA 94720, USA %9
\and
	Berkeley University of California, Berkeley, CA 94720-4206, USA %10
\and
	University of Oxford, Oxford OX1 2JD, United Kingdom %11
\and
	Instituto de Astrofísica de Canarias, La Laguna, Tenerife, Spain %12
\and 
    Departamento de Astrofísica, Universidad de La Laguna, Tenerife, Spain % 13
}

\abstract{
In this paper we present the European Low Frequency Survey (ELFS), a project that will enable foregrounds-free measurements of primordial $B$-mode polarization to a level 10$^{-3}$ by measuring the Galactic and extra-Galactic emissions in the 5--120\,GHz frequency window. Indeed, the main difficulty in measuring the B-mode polarization comes not just from its sheer faintness, but from the fact that many other objects in the Universe also emit polarized microwaves, which mask the faint CMB signal. The first stage of this project will be carried out in synergy with the Simons Array (SA) collaboration, installing a 5.5--11 GHz coherent receiver at the focus of one of the three 3.5\,m SA telescopes in Atacama, Chile (``ELFS on SA''). The receiver will be equipped with a fully digital back-end based on the latest Xilinx RF System-on-Chip devices that will provide frequency resolution of 1\,MHz across the whole observing band, allowing us to clean the scientific signal from unwanted radio frequency interference, particularly from low-Earth orbit satellite mega-constellations. This paper reviews the scientific motivation for ELFS and its instrumental characteristics, and provides an update on the development of ELFS on SA.
}

\maketitle
%

%For bibliography use \cite{Keruzore:2022tpj}
%Don't forget to give each section, subsection, subsubsection, and paragraph a unique label (see Sect.~\ref{sec-intro}).

\section{Introduction}
\label{sec-intro}

    The high-fidelity subtraction of galactic foreground emission may be the largest single challenge to detecting the signature of primordial gravitational waves that could be generated by inflation shortly after the Big Bang. The target sensitivity of $r\sim 10^{-3}$ in the next generation CMB experiments, such as Simons Observatory (SO) implies that foregrounds are at least an order of magnitude higher in brightness.

There has been a great focus on the Galactic dust foreground, especially after the claimed BICEP2 $B$-mode detection turned out to be emission from dust. The level of complexity of the dust emission is not yet fully known, and it is recognized by the CMB community that dust has to be very well understood to rule out its role in a future claim of $B$-modes detection. 
Synchrotron radiation is less well characterized and potentially more complex than dust. As measurements become deeper, synchrotron could become a bigger obstacle than dust if its spectrum turns out to be highly complex.

Our current knowledge of the synchrotron radiation has increased significantly after ground-based experiments operating in the 2--20\,GHz range: S-PASS
(2.3\,GHz) \cite{krachmalnicoff2018} in the Southern Hemisphere, C-BASS (5\,GHz) \cite{jones2018} and QUIJOTE-MFI (10--20\,GHz) \cite{MFIwidesurvey} in the Northern Hemisphere. These data highlight that the synchrotron emission is more complex than has been assumed in CMB forecast codes. Data from low-frequency instruments will be key in constraining and removing the synchrotron emission to extract the CMB signal.

The European Low Frequency Survey (ELFS) is a long-term plan to deploy dedicated telescopes to produce a full-sky survey in the 5--100\,GHz range with an unprecedented level of angular resolution ($\sim$20 arcmin at 10\,GHz), sub-GHz spectral resolution and sensitivity that will allow B-mode extraction from data produced by LiteBIRD, Simons Observatory and CMB-Stage 4 measurements. A first step (that we call ELFS on SA) is the deployment of a 6--12\,GHz  receiver (hereafter P6/12) in the Gregorian focus of one of the Simons Array telescopes \cite{barron2018}, followed by the installation of the QUIJOTE-MFI2 \cite{hoyland2022} to cover the 10--20\,GHz range. 

In this paper we briefly review the instrumental characteristic of the P6/12 receiver and show the impact of ELFS on SA measurements when combined with those expected from the Simons Observatory.

\section{ELFS on Simons Array: the P6/12 instrument}
\label{sec-instrument}

\begin{figure}[h!]
        \begin{center}
            \includegraphics[width=13cm]{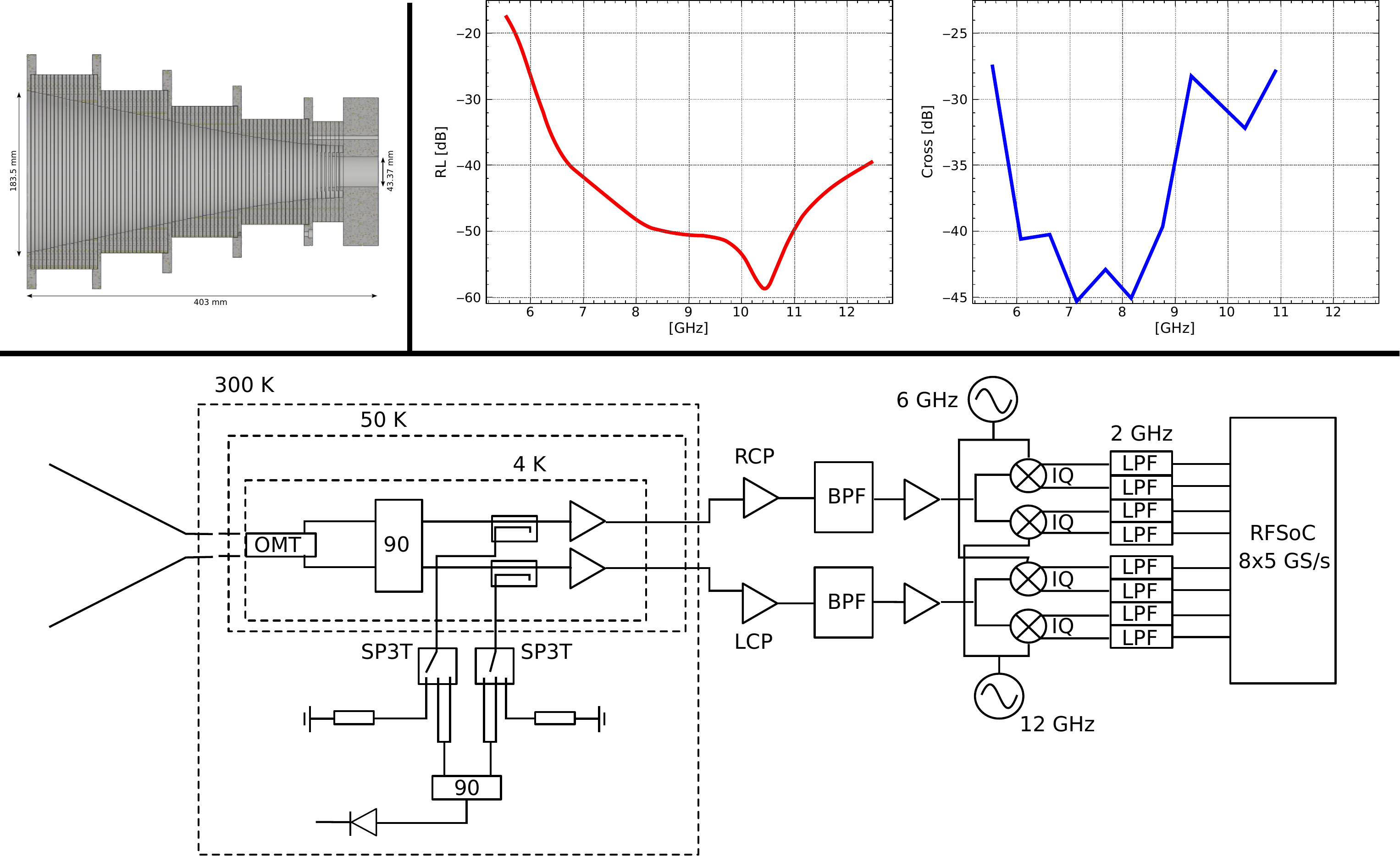}
        \end{center}
        \caption{\label{fig-instrument}\textit{Top panel}: electromagnetic model of the corrugated feedhorn (left), simulated return loss (center) and cross-polarization (right). \textit{Bottom panel}: receiver schematic.}
    \end{figure}

        In Table~\ref{tab-instrument-parameters} we report the main instrumental performance in terms of frequency bands, angular resolution and noise of the P6/12 receiver. A schematic of the main components is reported in Fig.~\ref{fig-instrument}.
%    \begin{mdframed}
%    \lipsum[22]
%    \end{mdframed}
    \begin{table}[h!]
    \begin{center}
        \caption{\label{tab-instrument-parameters} Top-level performanance parameters of the P6/12 receiver. $N^\text{P}$ indicates the noise level in polarization.}
        \begin{tabular}{c c c c c}
        \hline
        $\nu$\,[GHz] & $\theta_\text{FWHM}$\,[arcmin] &
        $N^\text{P}$\,[$\mu$K$\cdot$arcmin] &
        $\ell_\text{knee}$ & $\alpha_\text{knee}$\\
        \hline
        \phantom{1}6.3 & 46.6 & 539 & 15 & -2.4\\
        \phantom{1}7.0 & 42.2 & 512 & 15 & -2.4\\
        \phantom{1}7.7 & 38.1 & 487 & 15 & -2.4\\
        \phantom{1}8.6 & 34.4 & 465 & 15 & -2.4\\
        \phantom{1}9.5 & 31.1 & 443 & 15 & -2.4\\
        10.5& 28.1 & 423 & 15 & -2.4\\
        \hline
        \end{tabular}

    \end{center}
    \end{table}

% \begin{figure}[h!]
%     \begin{center}
%         \includegraphics[width=\linewidth]{figures/feedhorn_crosspol.jpeg}
%     \end{center}
%     \caption{Feedhorn cross polarization within the frequency band. It stays well below -35 dB in the central and left sides of the band and around -30 dB in the right side.}
%     \label{feed_cross}
% \end{figure}

The top-left panel of Fig.~\ref{fig-instrument} shows the feedhorn mechanical design with the main dimensions. Following the ideas presented in \cite{1487785} we have designed it with a hyperbolic profile and a ring-loaded slot mode converter. The corrugation teeth are 4.5\,mm in the body of the horn and 5\,mm in the mode converter part, with constant tooth/groove ratio of 0.889. This configuration allowed us to obtain excellent broadband performance in terms of return loss and cross-polarization performance, as shown in the top-center and top-right plots of Fig.~\ref{fig-instrument}. 

% \begin{figure}[h!]
%     \begin{center}
%         \includegraphics[width=\linewidth]{figures/feedhorn_mech.png}
%     \end{center}
%     \caption{Feedhorn mechanical design.}
%     \label{feed_mech}
% \end{figure}

% \begin{figure}[h!]
%     \begin{center}
%         \includegraphics[width=\linewidth]{figures/feedhorn_mech_sec.png}
%     \end{center}
%     \caption{Feedhorn mechanical design.}
%     \label{feed_mech_sec}
% \end{figure}

The orthomode transducer (OMT) is based on a design being used for the Square Kilometre Array Band 5a and 5b feeds. This is a quad-ridge design in which the ridges for one polarization are held at a constant spacing while they pass by the coaxial probe and backshort of the other polarization. This design reduces coupling between the polarizations and allows for a bandwidth of more than 2:1 with minimal excitation of higher order modes. The OMT is manufactured in four quadrants which are assembled with precision dowels to set the critical spacings between the ridges, which then requires no further tuning.  

The OMT will be incorporated into a cryostat adapted from the C-BASS North receiver \citep{2014MNRAS.438.2426K}. This is based on a Sumitomo SRDK-408D2 two-stage Gifford-McMahon cold head that reaches a base temperature of 4\,K. The cryostat cold plate can be interfaced up to four low-noise amplifiers plus the associated planar hybrid modules to allow for the continuous comparison radiometer architecture used in C-BASS. It will be modified to accommodate the new 2:1 bandwidth OMT in place of the 30\,\% bandwidth OMT used in C-BASS. 

As ELFS is focussed on polarization measurements for which the continuous comparison architecture is not necessary, we will use just two RF channels, using a planar 90-degree hybrid to convert the linear OMT outputs to circular polarization, and two Low Noise Factory LNF\_LNC4\_16C amplifiers to provide the primary gain (see bottom panel of Fig.~\ref{fig-instrument}). Coaxial 30\,dB couplers in the RF lines before the LNAs will allow us to inject a noise signal for calibration in to both circular polarizations.  
    
 %   \begin{mdframed}
  %   \lipsum[32]
  %  \end{mdframed}

%    \begin{mdframed}
%     \lipsum[28]
%    \end{mdframed}
    
%    \begin{mdframed}
%     \lipsum[8]
%    \end{mdframed}
    
%     \begin{figure}[h!]
%     \begin{center}
%         \includegraphics[width=10cm]{figures/beams.pdf}
%     \end{center}
%     \caption{\label{fig-beams}\textit{Left panel}: electromagnetic model of the dual-reflector, gregorian Simons Array telescope. \textit{Right panel}: electromagnetic models of the corrugated feedhorn (top) and of the OMT (bottom).}
% \end{figure}

The backend will follow the design philosophy of the QUIJOTE-MFI2 instrument \citep{hoyland2022}, using an FPGA based digital backend. The MFI2 FPGA (Xilinx ZCU208 Ultrascale) has the capability to simultaneously acquire eight RF channels at a sampling frequency of 5.0\,GSps, a $2.5$\,GHz band and 1\,MHz spectral resolution. The full bandwidth will be divided into  spectral sub-bands with maximum bandwidth of 2.5\,GHz, which are down-converted to base band $[0,2.5]$\,GHz through separate Local Oscillators (LOs). In the P6/12 design this will be achieved by using the complex output of mixers to achieve two bands from any LO. 

The backend design implements a hard programmed polyphase filterbank and a fast Fourier transform to retrieve the spectral information from the digitised samples of the fast onboard ADCs in the time domain. The power spectral density is then integrated in time and averaged. A temporary storage space is used to store 24\,h of raw data which is used to find an optimum blocking filter for any undesired interference. The final stored scientific signal is a spectral average with an averaging factor that depends on the scientific needs and the available storage space.

\section{Expected impact on science}
\label{sec-science}

\begin{figure}[h!]
    \begin{center}
        \includegraphics[width=11.2cm]{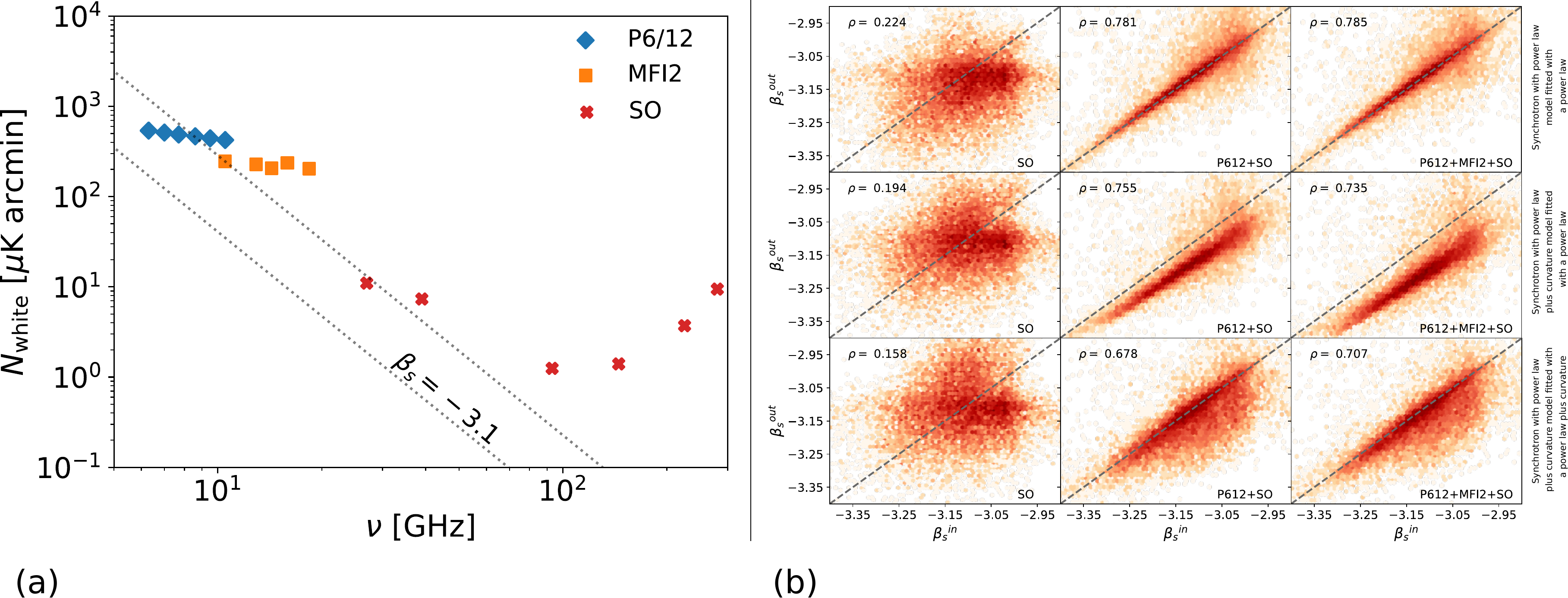}
    \end{center}
    \caption{\label{analysis_cases_foregrounds_fit}(a) Sensitivity of the tested instrument configurations as a function of frequency. The dotted line shows the extrapolated sensitivity following a power law with a $-3.1$ exponent.\\
    (b) Comparison of recovered $\beta_s$ values to input $\beta_s$ values for all observable pixels from Atacama. From left to right: SO, P612+SO, and P612+MFI2+SO results. The dashed diagonal represents $\beta_s^{\rm out} = \beta_s^{\rm in}$. The correlation coefficients ($\rho$) are displayed in the upper left of each plot.}
\end{figure}

    To assess the impact of adding low frequency channels in foreground and CMB data analysis, we considered the Simons Observatory\footnote{We include the additional SATs from SO:UK and SO:JP as well as an extended observation time (until 2035).} (SO) experiment combined with ELFS-SA channels (the P6/12 and the QUIJOTE-MFI2 instruments), as shown in panel (a) of Fig.~\ref{analysis_cases_foregrounds_fit}, which shows frequencies and sensitivities. 

Our simulations include a sky signal with the CMB, synchrotron emission with or without curvature (i.e., frequency-dependent spectral index), and thermal dust emission. We used \texttt{d10} and \texttt{s5} PySM models \cite{pysm} to simulate the thermal dust and the synchrotron emission without curvature. To simulate the synchrotron with a curved spectrum we generated a custom model with the same spectral index as \texttt{s5} and a curvature template whose values are allowed by current observations in the Southern Hemisphere. This template is created by mean shifting to achieve a final mean of 0.04 and rescaling the spatial variations by a factor of $\sim 30$ the PySM \texttt{s7} template. We emphasize that neither the sky model nor the cleaning procedure was validated through the SO pipeline presented in Wolz et al. \cite{wolz2023}.

In the right panel of Fig.~\ref{analysis_cases_foregrounds_fit}, we display the fitted synchrotron spectral index for various instrument configurations. It shows that, for the synchrotron model considered, a good characterization is reached by combining ELFS with SO. The first row shows the case where synchrotron follows a power-law and is fitted with a power-law. The second row represents the scenario where synchrotron has a curved spectral index but is fitted with a power-law. Lastly, the third row demonstrates the case where curved synchrotron is fitted with the correct model. The bias observed in the second row of fits with ELFS confirms that including low-frequency channels allows us to detect synchrotron curvature.

\begin{figure}[h!]
    \begin{center}
        \includegraphics[width=11.2cm]{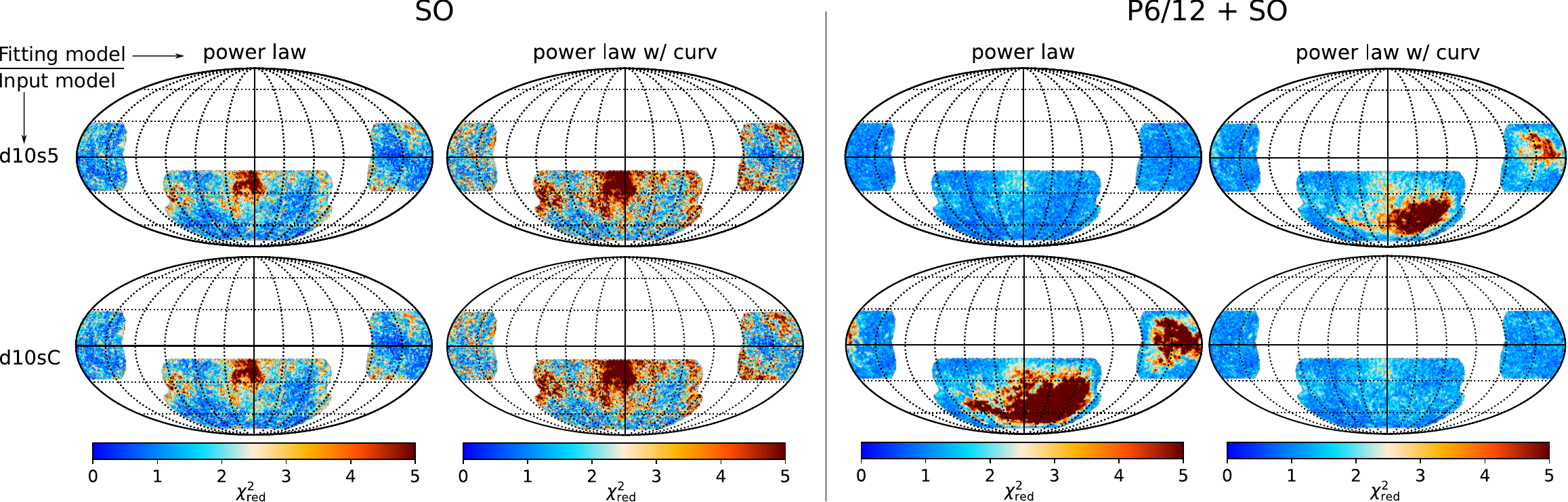}
        \caption{\label{fig_chisq}Reduced $\chi^2$ maps generated using SO data alone and ELFS+SO data. The top (bottom) row corresponds to simulations using the \texttt{d10s5} (\texttt{d10sC}) model, where synchrotron is represented as a power law (power law with curvature). The odd columns display the maps obtained with a power law model fitting the synchrotron emission, while the even columns present the $\chi^2_{\rm red}$ values when the synchrotron is modeled with a power law with curvature.}
    \end{center}
\end{figure}

Figure~\ref{fig_chisq} shows the reduced $\chi^2$ from the synchrotron fits obtained using SO alone (left) or ELFS-SA plus SO (right). The comparison of the top and bottom rows shows how the existence of an extra parameter in synchrotron complexity may be effectively constrained by the combination of ELFS and SO data. When we include the information from ELFS on SA, we observe differences between the two simulated skies, obtaining the best fit when we fit the synchrotron using the correct model.

\begin{figure}[h!]
    \begin{center}
    \includegraphics[width=10.2cm]{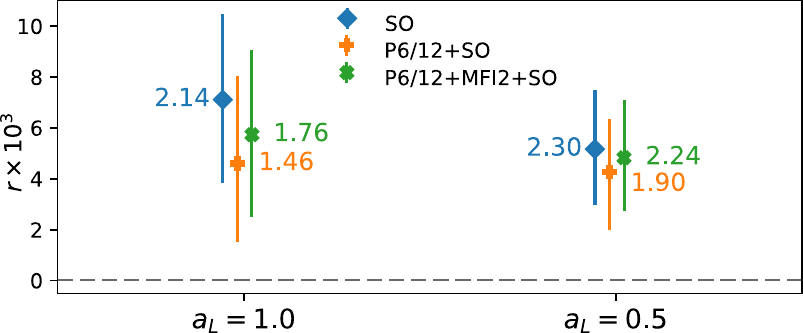}
    \end{center}
    \caption{\label{fig_r_constraint}The tensor-to-scalar ratio recovery, simulated with \texttt{d10sC} foregrounds, varies with the amount of included lensing in the CMB simulations. Blue diamonds, orange plus signs, and green crosses represent values obtained using the best-fit model for SO, P6/12+SO, and P6/12+MFI2+SO data, respectively. This involves using a power-law model for SO alone and a power-law with curvature when including low-frequency data.}
\end{figure}

Figure~\ref{fig_r_constraint} shows the recovered tensor-to-scalar-ratio when the sky is modeled using the \texttt{d10sC} foreground models, in two cases: without delensing, and assuming a 50\% delensing. The $r$ value shown corresponds to the value obtained with the best-fit model from the $\chi^2_{\rm red}$ maps, i.e., the power-law model in the case of SO alone, and power-law with curvature when we add the instruments from ELFS on SA. Results indicate how the bias on $r$ can be mitigated when ELFS and SO data are combined. The value of $r$ remains somewhat biased due to the challenge of accurately determining the true combination of $\beta_s$ and $c_s$ values, which are strongly degenerate. We are currently exploring approaches to mitigate this bias, including the incorporation of prior information related to the effective exponent in the power-law with curvature model to break the parameter degeneracy and enhancing the sensitivity of the P6/12.

These results highlight the importance of low-frequency data in discerning complex synchrotron behavior, extending beyond a basic power-law model. Furthermore, in the latter case, these data will help mitigate the residuals coming from the synchrotron emission.

\section{Conclusions}
\label{sec-conclusions}

    In this paper we have presented ELFS, a plan to deploy dedicated telescopes to produce a full-sky survey in the 5--100\,GHz range, and its first incarnation, ELFS on SA, which foresees the deployment of a 6--12\,GHz receiver in the Gregorian focus of one of the Simons Array telescopes, followed by the installation of the QUIJOTE-MFI2 to cover the 10--20\,GHz range. 
We have analyzed the potential of ELFS on SA in discerning complex synchrotron behavior, when combined with measurements from next generation experiments, like the Simons Observatory. Our results show that the complexity of synchrotron cannot be underestimated and the availabilty of high precision, low frequency observations will be of key importance in presence of a potential tensor-to-scalar ratio detection.

\section*{Acknowledgements}

This is not an official SO Collaboration paper.

\bibliography{biblio}

\end{document}